\documentstyle[12pt]{article}

\renewcommand{\thefootnote}{\fnsymbol{footnote}}
\newcommand{\EQ}{\begin{equation}}
\newcommand{\EN}{\end{equation}}
\newcommand{\bea}{\begin{eqnarray}}
\newcommand{\ena}{\end{eqnarray}}
\newcommand{\vs}[1]{\vspace{#1 mm}}
\newcommand{\hs}[1]{\hspace{#1 mm}}

\renewcommand{\b}{\beta}

\newcommand{\pa}{\partial}

\newcommand{\uda}{\nearrow \kern-1em \searrow}

\begin{document}

\topmargin 0pt
\oddsidemargin 5mm

\begin{titlepage}
\setcounter{page}{0}
\begin{flushright}
OU-HET  \\
June, 1997
\end{flushright}

\vs{12}
\begin{center}
{\Large  Double Beta Decay Constraint on Composite Neutrinos}
\vs{15}

{\large 
Eiichi Takasugi\footnote{e-mail address: takasugi@phys.wani.osaka-u.ac.jp}}\\
\vs{8}
{\em Department of Physics, \\
Osaka University \\ Toyonaka, Osaka 560, Japan} \\
\end{center}
\vs{6}

\centerline{{\bf{Abstract}}}

Neutrinoless double beta decay $(\b\b)_{0\nu}$ occurs 
through the magnetic 
coupling of dimension five, $\lambda_W/m_{\nu*}$,  among the excited 
electron neutrino $\nu^*$, electron and $W$ boson if $\nu^*$  is a massive 
Majorana neutrino. 
\bea
\left( \frac{\lambda_W m_W}{m_{\nu*}}
\right)^2
\left | \frac{\frac{m_*}{m_W}+2}{(\frac{m_*}{m_W}+1)^2}
 -\frac{0.129}{\frac{m_*}{m_W}}\right | <2.13\times 10^{-2}\;,
\nonumber
\ena
where $\lambda_W$ is the relative strength, $m_{\nu*}$ is the 
composite scale and $m_*$ is the mass of excited neutrino. 
If the coupling  is not small, i.e., $\lambda_W>1$ and 
$m_{\nu*}=m_*$,  we find $m_*>3.4 m_W$ which is the most stringent 
limit.  

\end{titlepage}
\newpage
\renewcommand{\thefootnote}{\arabic{footnote}}
\setcounter{footnote}{0}
\section{Introduction}

If neutrinos are composite particles, there exist  the excited neutrinos 
which couple to the ground state leptons by the dimension five 
magnetic coupling[1],[2].  This interaction is 
expressed as[3]
\bea
L_{int}=g\frac{\lambda_W }{ m_{\nu^*}}\bar e 
\sigma^{\mu\nu}(\eta_L^* R+\eta_R^* L)\nu^* \pa_\mu W_\nu^-  + h.c. ,
\ena
where $\nu^*$ is a heavy excited electron neutrino, $L=(1-\gamma_5)/2$, 
$R=(1+\gamma_5)/2$, $m_{\nu*}$ is the mass dimension which is of 
order the mass of $\nu^*$, i.e., $m_*$. This interaction is derived by the 
$SU(2)\times U(1)$gauge invariance  and parameters $\eta_L$ 
and  $\eta_R$ are normalized by   
\bea
\mid \eta_L \mid ^2 + \mid \eta_R \mid^2 =1\;. 
\ena
The extensive search of $\nu^*$ have been made by many 
groups[4] and found that $m_{\nu*} > 91$GeV by assuming 
that  $\lambda_Z > 1$ which is the coupling for 
$\nu^* \to e Z$ decay similarly defined  to  $\lambda_W $. 

The purpose of this paper is to explore the mass range 
$m_{\nu*} > m_Z$ by using the neutrinoless double beta 
decay $(\b\b)_{0\nu}$  by assuming that  $\nu^*$ is  a 
massive Majorana neutrino.  Then the $(\b\b)_{0\nu}$ decay 
 occurs through $\nu^*$  exchange.   Since $\nu^*$  enters 
 as a virtual state,  we can investigate heavy $\nu^*$.  

In our previous paper[5], we started from the interaction 
in Eq.(1) and derived  the 
effective four point interaction between leptons and hadrons as
\bea
L_{eff}=-G_{eff}\bar e \sigma^{\mu\nu}(\eta_L^* R+
\eta_R^* L)\nu^* \pa_\mu J_\nu^\dagger+ h.c. ,
\ena
where $J_\nu^\dagger$ is the hadronic current and 
\bea
G_{eff}=2G_F \frac{g\lambda_W^{(\nu^*)} }{m_{\nu^*}}.
\ena
From this effective interaction with the chirality selection 
rule $\eta_L \eta_R=0$, the 
half-life of the neutrinoless double beta decay was derived[5] 
and is given by 
\bea
T^{-1}= 4\left( \frac{\lambda m_A}{m_{\nu*}}\right)^4 
\left(\frac{m_A}{m_e}\right)^2 G_{01}
\left | \frac{m_*}{m_A}
)M_{GT,N}+\frac{m_A}{m_*}[(\frac{g_V}{g_A})^2 M_F'
-\frac 23 M_{GT}'-\frac 13 M_T']\right | ^2 
 \;,
\ena
which disagreed with the old result by Panella and Srivastava[6]. 
By comparing this formula to the Heidelberg-Moscow data[7] 
for ${}^{76}$Ge decay, we found[6] 
\bea
\left( \frac{\lambda m_A}{m_{\nu*}}\right)^2
\mid \frac{m_*}{m_A} +
7.2\cdot 10^{-2}\frac{m_A}{m_*}\mid < 1.4\cdot 10^{-8}.
\ena
The formula in Eq.(5) and also the constraint in Eq.(6) 
are valid as far as we consider the effective interaction 
in Eq.(3). 
 
Recently, Panella, Carimalo, Srivastava and Widom pointed out[8] 
that  propagators of W-boson exchange 
should not be contracted as far as one sticks to the interaction 
in Eq.(1). This is true since the mass of excited neutrinos 
$m_*$ is  greater than $m_W$. They calculated[8] the half-life 
formula of the neutrinoless double beta decay and claimed that 
my calculation is wrong. Here, I say that my calculation is 
correct as far as we consider  the effective interaction in 
Eq.(3). Anyway, I feel to be obliged to repeat the computation 
starting from the interaction in Eq.(1). 
 
In Sec. 2, we show the computation of the decay and give 
the half-life formula. In Sec. 3, the numerical analysis will 
be given. Summary is presented in Sec.4.

\section{Decay formula of neutrinoless double beta decay}

In the fourth order perturbations of the  interaction 
in Eq.(1), the $(\b\b)_{0\nu}$ decay  takes place and the S-matrix 
for this decay is given by 
\bea
S&=&-i{{G_{eff}^2m_W^4} \over{2 (2\pi)^{12}}} 
\int dx_1dx_2dx_3dx_4
dqdq_1dq_2
{{ e^{-iq(x_1-x_2)}} \over {q^2-m^2_*}}
{{ e^{-iq_1(x_1-x_3)}} \over {q_1^2-m^2_W}}
{{ e^{-iq_2(x_2-x_4)}} \over {q_2^2-m^2_W}}q_{1\mu}q_{2\mu'}
\nonumber\\
&&\times 
T(\bar e(x_1) \sigma^{\mu\nu}\sigma^{\mu'\nu'}
[m_*(\eta_L^{*2 }R+\eta_R^{*2} L)
+\gamma_{\mu}q^{\mu}\eta_L^{* }
\eta_R^{* }]
 e^C(x_2))T(J_\nu^{\dagger}(x_3) J_{\nu'}^{\dagger}(x_2))\; ,
\nonumber\\
\ena
where $e^C$ is the charge conjugation of $e$, i.e., $e^C=C\bar e^T$. 

In the following computation, we take  the S-wave function for 
electrons which is given by
\bea
<0\mid e(x) \mid p>= \psi_S(\epsilon) e^{-i\epsilon x^0};
\hs{3}
\psi_S(\epsilon)=\sqrt{{\epsilon+m} \over {2\epsilon}}
\left(\begin{array}{c}
\chi_s \\
\frac{\vec \sigma\cdot \vec p}{\epsilon+m} \chi_s
\end{array} \right)
F_0(Z,\epsilon) ,
\ena
where $\epsilon$ is the energy of electron and $F_0(Z,\epsilon) $ is the 
relativistic Coulomb factor defined in Eq.(3.1.25) in Ref.9. The S-wave 
function  is independent of the space coordinate. After integrating 
$x_1$, $x_2$ and then $q_1$, $q_2$,  we find 
\bea
S_{fi}&=&-i\frac{G_{eff}^2m_W^4}{2 (2\pi)^{12}}
 \int dx_3 dx_4 \int dq
\frac{ e^{-iq(x_3-x_4)}} {q^2-m^2_*}
<N_f|T(J_\nu^{\dagger}(x_3) 
J_{\nu'}^{\dagger}(x_4))|N_i>\nonumber\\
&&\times \frac{e^{i(\epsilon_1x_3^0+\epsilon_2x_4^0}} 
{((-q+k_1)^2-m^2_W)((q+k_2)^2-m^2_W)}
(-q+k_1)_{\mu}(q+k_2)_{\mu'}
\nonumber\\
&&\times 
\bar \psi_S(\epsilon_1) \sigma^{\mu\nu}\sigma^{\mu'\nu'}
[m_*(\eta_L^{*2 }R+\eta_R^{*2} L)
+\gamma_{\mu}q^{\mu}\eta_L^{* }\eta_R^{* }]
\psi_S^C(\epsilon_2)
\nonumber\\
&&\hskip 1cm -(\epsilon_1\iff \epsilon_2)\; ,
\nonumber\\
\ena
where $k_i=(\epsilon_i,0)$. 

Next we perform the $q^0$ integration. There are six poles at 
 $\pm (E_* -i\epsilon)$, $\epsilon_1 \pm (E_W-i\epsilon)$ and 
$-\epsilon_2 \pm (E_W-i\epsilon)$. 
The poles in the lower-half plain contribute when 
$x_3^0-x_4^0>0$ and those in the upper-half plain do when 
$x_3^0-x_4^0<0$. By performing the $q^0$ integration, 
the time ordering 
for hadronic currents is dealt automatically. We find 
 
\bea
S_{fi}&=&-\frac{G_{eff}^2m_W^4}{4 (2\pi)^{3}}
 \int dx_3 dx_4 \int d\vec{q}
 e^{i\vec{q}\cdot(\vec{x_3}-\vec{x_4})} 
e^{i(\epsilon_1 x_3^0+\epsilon_2 x_4^0)}
\nonumber\\ 
&&\times \bar \psi_S(\epsilon_1)
 \sigma^{\mu\nu}\sigma^{\mu'\nu'}
[m_*(\eta_L^{*2 }R+\eta_R^{*2} L)
+\gamma_{\mu}q^{\mu}\eta_L^{* }\eta_R^{* }]
\psi_S^C(\epsilon_2)
\nonumber\\
&&\times 
  [\theta (x_3^0-x_4^0)<N_f|J_\nu^{\dagger}(x_3)|n>
  <n|J_{\nu'}^{\dagger}(x_4)|N_i>A_{\mu\mu'}
\nonumber\\
&&\hskip 2mm +
\theta (x_4^0-x_3^0)<N_f|J_{\nu'}^{\dagger}(x_4) |n>
<n|J_\nu^{\dagger}(x_3) |N_i>B_{\mu\mu'}]
\nonumber\\
&&
 \hskip 1cm -(\epsilon_1\iff \epsilon_2)\; ,
\ena
where
\bea
A_{\mu\mu'}&=&\frac{(q_*-k_1)_{\mu}
(q_*+k_2)_{\mu'}e^{-iE_*(x_3^0-x_4^0)}}
   {E_*[(E_*-\epsilon_1)^2-E_W^2]*[(E_*+\epsilon_2)^2-E_W^2]}
\nonumber\\
&&+\frac{q_{W\mu}(q_W+k_1+k_2)_{\mu'}
e^{-i(E_W+\epsilon_1)(x_3^0-x_4^0)}}
   {E_W[(E_W+\epsilon_1)^2-E_*^2]
[(E_W+\epsilon_1+\epsilon_2)^2-E_W^2]}
\nonumber\\
&&+\frac{(q_W-k_1-k_2)_{\mu}q_{W\mu'}
e^{-i(E_W-\epsilon_2)(x_3^0-x_4^0)}}
   {E_W[(E_W-\epsilon_2)^2-E_*^2]
[(E_W-\epsilon_1-\epsilon_2)^2-E_W^2]}\;,
\nonumber\\
B_{\mu\mu'}&=&\frac{(\tilde{q}_*+k_1)_{\mu}(\tilde{q}_*-k_2)_{\mu'}
e^{iE_*(x_3^0-x_4^0)}}
   {E_*[(E_*+\epsilon_1)^2-E_W^2][(E_*-\epsilon_2)^2-E_W^2]}
\nonumber\\
&&+\frac{(\tilde{q}_W+k_1+k_2)_{\mu}\tilde{q}_{W\mu'}
e^{i(E_W+\epsilon_2)(x_3^0-x_4^0)}}
   {E_W[(E_W+\epsilon_2)^2-E_*^2]
[(E_W+\epsilon_1+\epsilon_2)^2-E_W^2]}
\nonumber\\
&&+\frac{\tilde{q}_{W\mu}(\tilde{q}_W-k_1-k_2)_{\mu'}
   e^{i(E_W-\epsilon_1)(x_3^0-x_4^0)}}
   {E_W[(E_W-\epsilon_1)^2-E_*^2]
[(E_W-\epsilon_1-\epsilon_2)^2-E_W^2]}\;.
\ena
Here $E_*=\sqrt{m_*^2+\vec q^2}$, $q_*=(E_*,\vec q)$ and 
$\tilde q_*=(E_*,-\vec q)$ and similarly for $E_W$, $q_W$ and 
$\tilde q_W$. 
Then, we perform the $x_3^0$ and $x_4^0$ integration. The 
$\vec x_3$ and $\vec x_4$ integration can be made immediately 
since we use the non-relativistic approximation of hadronic current,
\bea
J_\mu^\dagger(\vec x)=\sum_j J_\mu^\dagger (j)
 \delta (\vec x-\vec r_j),\hs{2}
J_\mu^\dagger (j) =\tau_j^+ (g_V g_{\mu 0}+g_A \delta_{\mu l} 
\sigma_j^l)F(q^2)\;,
\ena
where $\vec r_j$ is the position of the j-th nucleon in the nucleus and  
$F(q^2)$ is the form factor defined by
\bea
F(q^2)=(\frac{1}{1+(q^2/m_A^2)})^2,
\ena
with the value $m_A=0.85$GeV.  By these integration, we obtain the 
energy conservation $E_i=E_f+\epsilon_1+\epsilon_2$ and the 
energy denominators.  
We find 
\bea
R_{fi}&=&\frac{G_{eff}^2m_W^4}{\sqrt{2!} 4}
\bar \psi_S(\epsilon_1) \sigma^{\mu\nu}\sigma^{\mu'\nu'}
[m_*(\eta_L^{*2 }R+\eta_R^{*2} L)+\gamma_{\mu}q^{\mu}
\eta_L^{* }\eta_R^{* }]
\psi_S^C(\epsilon_2)
\nonumber\\
&&\times 
\sum_{j\neq l}\int \frac{d\vec{q}}{(2\pi)^3}e^{-\vec q \cdot\vec r_{jl}}
\sum_n
[<N_f|J_\nu^{\dagger}(\vec r_j) |n>
<n|J_{\nu'}^{\dagger}(\vec r_l) |N_i>C_{\mu\mu'}
\nonumber\\
&&
\hskip 2cm +<N_f|J_{\nu'}^{\dagger}(\vec r_l)|n>
<n|J_\nu^{\dagger}(\vec r_j)|N_i>D_{\mu\mu'}]
\nonumber\\
&&
 \hskip 2cm -(\epsilon_1\iff \epsilon_2)\; ,
\ena
where $\epsilon_n=E_n-(E_i+E_f)/2$ and 
\bea
C_{\mu\mu'}&=&\frac{(q_*-k_1)_{\mu}(q_*+k_2)_{\mu'} }
   {(E_*+\epsilon_n-\frac{\epsilon_1-\epsilon_2}{2})
   E_*[(E_*-\epsilon_1)^2-E_W^2][(E_*+\epsilon_2)^2-E_W^2]}
\nonumber\\
&&+\frac{q_{W\mu}(q_W+k_1+k_2)_{\mu'} }
   {(E_W+\epsilon_n+\frac{\epsilon_1+\epsilon_2}{2})
   E_W[(E_W+\epsilon_1)^2-E_*^2]
[(E_W+\epsilon_1+\epsilon_2)^2-E_W^2]}
\nonumber\\
&&+\frac{(q_W-k_1-k_2)_{\mu}q_{W\mu'} }
   {(E_W+\epsilon_n-\frac{\epsilon_1+\epsilon_2}{2})
   E_W[(E_W-\epsilon_2)^2-E_*^2]
[(E_W-\epsilon_1-\epsilon_2)^2-E_W^2]}
\nonumber\\
D_{\mu\mu'}&=&\frac{(\tilde{q}_*+k_1)_{\mu}
(\tilde{q}_*-k_2)_{\mu'}}
   {(E_*+\epsilon_n+\frac{\epsilon_1-\epsilon_2}{2})
   E_*[(E_*+\epsilon_1)^2-E_W^2][(E_*-\epsilon_2)^2-E_W^2]}
\nonumber\\
&&+\frac{(\tilde{q}_W+k_1+k_2)_{\mu}\tilde{q}_{W\mu'} }
   {(E_W+\epsilon_n+\frac{\epsilon_1+\epsilon_2}{2})
   E_W[(E_W+\epsilon_2)^2-E_*^2]
[(E_W+\epsilon_1+\epsilon_2)^2-E_W^2]}
\nonumber\\
&&+\frac{\tilde{q}_{W\mu}(\tilde{q}_W-k_1-k_2)_{\mu'}}
   {(E_W+\epsilon_n-\frac{\epsilon_1+\epsilon_2)}{2})
   E_W[(E_W-\epsilon_1)^2-E_*^2]
[(E_W-\epsilon_1-\epsilon_2)^2-E_W^2]}
\nonumber\\
\ena

Hereafter, we consider the case where  the chirality selection rule 
is satisfied, i.e., $\eta_L\eta_R=0$ . 
So far we took (a1) the S-wave function for electron wave function and 
thus the total angular momentum taken by electrons are 0 or 1 so that 
the $0^+\to 0^+$ and $0^+\to 1^+$ transitions are allowed in general. 
Then, we used (a2) the non-relativistic 
approximation of the hadronic current because we are dealing with the 
allowed transition. Next, we make (a3) the closure approximation where 
 $\epsilon_n$ is replaced by the average value 
$<\epsilon_n>=\mu_0 m_e$. Then, the sum of the intermediate states can 
be taken. Then, we find 
\bea
\sum_n<N_f|J_\nu^{\dagger}(\vec r_j)|n>
<n|J_{\nu'}^{\dagger}(\vec r_l)|N_i>&=&
\sum_n<N_f|J_{\nu'}^{\dagger}(\vec r_l)|n>
<n|J_\nu^{\dagger}(\vec r_j)|N_i>
\nonumber\\
&=&<N_f|J_\nu^{\dagger}(\vec r_j)J_{\nu'}^{\dagger}(\vec r_l))|N_i>
\ena
 so that $C_{\mu\mu'}$ and $D_{\mu\mu'}$ enter in 
$S_{fi}$ as their sum. 

Now, we  concentrate on the $0^+ \to 0^+$ transition. Since 
$J_{\nu}(\vec r_j)$ is an parity even operator, only the 
$mu=\mu'=0$ and $\mu=k, \mu'=k'$ ($k, k'=1,2,3$) parts contribute to the 
$0^+ \to 0^+$ transition because $\vec q$ is an odd parity 
operator. Now, we find 
that $ C_{00}+D_{00}$ and $ C_{kk'}+D_{kk'}$ are even functions 
with respect to 
the exchange of $\epsilon_1$ and $\epsilon_2$. 
Then, the first part and the $(\epsilon_1\iff \epsilon_2)$ part are 
combined as
\bea
R_{fi}&=&\frac{G_{eff}^2m_W^4}{4\sqrt{2} }
\bar \psi_S(\epsilon_1) \{\sigma^{\mu\nu},\sigma^{\mu'\nu'}\}
m_*(\eta_L^{*2 }R+\eta_R^{*2} L)\psi_S^C(\epsilon_2)
\int \frac{d\vec{q}}{(2\pi)^3}e^{i\vec q \cdot\vec r_{jl}} 
\nonumber\\
&& \times 
\sum_{j\neq l}
 <N_f|J_\nu^{\dagger}(\vec r_j)J_{\nu'}^{\dagger}(\vec r_l)|N_i>
(C_{\mu\mu'}+D_{\mu\mu'})
\nonumber\\
\ena
Now we use the identity $\{\sigma^{\mu\nu},\sigma^{\mu'\nu'}\}
=2(g^{\mu\mu'}g^{\nu\nu'}-g^{\mu\nu'}g^{\nu\mu'}
-i\epsilon^{\mu\nu\mu'\nu'}\gamma_5)$.  Then, we find 
\bea
R_{fi}&=&-\frac{G_{eff}^2m_W^4m_*}{2\sqrt{2} }
\bar \psi_S(\epsilon_1)  
(\eta_L^{*2 }R+\eta_R^{*2} L)\psi_S^C(\epsilon_2)
\int \frac{d\vec{q}}{(2\pi)^3}e^{-\vec q \cdot\vec r_{jl}}
\nonumber\\
&&  \times 
 \sum_{j\neq l}
<N_f|(C_{00}+D_{00})\vec{J}^{\dagger} (\vec r_j)\cdot 
\vec{J}^{\dagger}
(\vec r_l)
\nonumber\\
&&\hskip 1cm 
+ (C+D)[\vec q^2(J_0^{\dagger}(\vec r_j)J_{0}^{\dagger}(\vec r_l)
 -\vec{J}^{\dagger} (\vec r_j)\cdot 
\vec{J}^{\dagger})+\vec q \cdot \vec{J}^{\dagger} (\vec r_j)
\vec q \cdot \vec{J}^{\dagger}
]|N_i>\;,
\nonumber\\
\ena
where we used $C_{kk'}+D_{kk'}=q_kq_{k'}(C+D)$. 

Now, we expand $C_{\mu\mu'}$ and $D_{\mu\mu'}$ with respect to 
the small quantities $\epsilon_i/E_*$,  $\epsilon_i/E_W$, 
$\epsilon_i/(m_*^2-m_W^2)$ and take the leading order 
terms. Here we assume $m_* > m_W$ and used the fact that 
  $\epsilon_i \simeq$ a few MeV. 
Then, we obtain
\bea
C_{00}+D_{00}&\simeq&- \frac{\mu_0 m_e
(E_*+2E_W)}{E_*E_W^3(E_*+E_W)^2}
\simeq - \frac{\mu_0 m_e(m_*+2m_W)}{m_*m_W^3(m_*+m_W)^2}
 \;,
\nonumber\\
C+D&\simeq&\frac{2}{E_*^2 E_W^4}\simeq \frac{2}{m_*^2 m_W^4}\;.
\ena
Here we used the fact that the momentum $q$ is effectively 
cut off by $m_A$ due to the form factor $F(q^2)$. 

Then, we use 
\bea
\int \frac{d \vec q}{(2\pi)^3}\frac{e^{i\vec q\cdot \vec r}}
  {(\vec q^2+m_A^2)^4}
  &=&\frac{1}{4\pi m_A^6}\frac{1}{r}F_N\;,
 \nonumber\\
\int \frac{d \vec q}{q_kq_l(2\pi)^3}\frac{e^{i\vec q\cdot \vec r}}
  {(\vec q^2+m_A^2)^4}
  &=&\frac{1}{4\pi m_A^4}\frac{1}{r}
  \frac 13 \left[  \delta_{kl}F_4 -
  (3\hat x_k \hat x_l -\delta_{kl})F_5
  \right ]\;,
\ena
where  with $x_A-m_A r$ 
\bea
F_N&=&\frac{x_A}{48}(3+3x_A+x_A^2)e^{-x_A},
\nonumber\\
F_4&=&\frac{x_A}{48}(3+3x_A-x_A^2)e^{-x_A},\nonumber \\
F_5&=&\frac{x_A^3}{48}e^{-x_A}
\ena
Then, we find  
\bea
R_{fi}&=&{{(G_{eff} m_A g_A)^2} \over{ 4\pi \sqrt{2}R}} 
\psi_S(\epsilon_1)
(\eta_L^{*2 }R+\eta_R^{*2} L)\psi_S^C(\epsilon_2)
F_0(Z+2,\epsilon_1)F_0(Z+2,\epsilon_2)\nonumber\\
&& \times
\left \{ \frac{\mu_0 m_e}2 \frac{m_W(m_*+2m_W)}{(m_*+m_W)^2}
M_{GT,N} -\frac{m_A^2}{m_*}[(\frac{g_V}{g_A})^2 M_F'
-\frac 23     M_{GT}'-\frac 13 M_T']\right\}.
\ena
where
\bea
M_{GT,N}&=&<N_f\mid\sum_{n\ne m}\tau_n^{(+)}\tau_m^{(+)} 
\vec \sigma_n\cdot\vec\sigma_m (\frac {R}{r_{nm}})
F_N(x_A)\mid N_i>,\\
M_{F}'&=&<N_f\mid\sum_{n\ne m}\tau_n^{(+)}\tau_m^{(+)} 
(\frac {R}{r_{nm}})
F_4(x_A)\mid N_i>,\\
M_{GT}'&=&<N_f\mid\sum_{n\ne m}\tau_n^{(+)}\tau_m^{(+)} \vec 
\sigma_n\cdot\vec\sigma_m (\frac {R}{r_{nm}})F_4(x_A)\mid N_i>,
\ena
\bea
M_T'=<N_f \mid\sum_{n\ne m}\tau_n^{(+)}\tau_m^{(+)} 
\{ 3(\vec \sigma_n\cdot\vec r_{nm})  
(\vec \sigma_m\cdot\vec r_{nm})
 - \vec\sigma_n\cdot\vec\sigma_m \}
 (\frac {R}{r_{nm}})F_5(x_A)\mid N_i>,
\ena

After taking the spin sum and performing the phase-space 
integration, we 
find the half-life of the  transition of the neutrinoless 
double beta decay due to the heavy composite neutrino for 
$0^+ \to 0^+$ transition as 
\bea
T^{-1}&=&4\left( \frac{\lambda m_A}{m_{\nu*}}\right)^4 
\left(\frac{m_A}{m_e}\right)^2 G_{01}
\left | ( \frac 12 \frac{\mu_0 m_e}{m_A}
\frac{m_W(m_*+2m_W)}{(m_*+m_W)^2}
)M_{GT,N}\right.\nonumber\\ 
&&\left.\hskip 1cm  -\frac{m_A}{m_*}[(\frac{g_V}{g_A})^2 M_F'
-\frac 23 M_{GT}'-\frac 13 M_T']\right |^2 ,
\ena
where $G_{01}$ is the phase space factor defined in Eq.(3.5.17a) 
in the paper by Doi, Kotani and Takasugi[9].  
 The above expression  is obtained by taking 
 $\mid \eta_L\mid^4+\mid\eta_R\mid^4=1$ which 
 is valid   because  of the chirality conservation.

\section{Constraint from neutrinoless double beta decay}

In the following, we analyze the constraint on the coupling by using 
the experimental half-life limit of  ${}^{76}{\rm Ge}$ which is 
measured by the Heidelberg-Moscow collaboration\cite{HM},
\bea
T(0^+ \to 0^+ :{}^{76}{\rm Ge}) > 5.6\cdot10^{24}yr \hs{3} 90\% c.l. . 
\ena
We derive the constraint on composite parameters from this data. 
By using the values of nuclear matrix elements obtained by Hirsch, 
Klapdor-Kleingrothaus and Kovalenko[10],
\bea
M_{GT,N}=0.113, \hs{1} M_{F}'=3.06\cdot10^{-3}, \hs{1} 
M_{GT}'=-7.70\cdot10^{-3}, \hs{1} M_T'=-3.09\cdot10^{-3},
\ena
 the phase space factor $G_{01}=6.4\cdot10^{-15}/yr$ 
given in Ref.6 and $g_A/g_V=1.25$, 
we find the half-life of the $0^+ \to 0^+$ transition for 
${}^{76}{\rm Ge}$ is 
\bea
T^{-1} =3.9\times 10^{-22} \left( \frac{\lambda_W 
m_W}{m_{\nu*}}
\right)^4 
\left | \frac{\frac{m_*}{m_W}+2}{(\frac{m_*}{m_W}+1)^2}
 -\frac{0.129}{\frac{m_*}{m_W}} \right |^2 ,
\ena
Then, we find the constraint
\bea
\left( \frac{\lambda_W m_W}{m_{\nu*}}
\right)^2
\left | \frac{\frac{m_*}{m_W}+2}{(\frac{m_*}{m_W}+1)^2}
 -\frac{0.129}{\frac{m_*}{m_W}} \right | <2.13\times 10^{-2}\;. 
\ena

\section{Discussions}

Firstly, we shall discuss about the excited neutrino mass. 
We consider the case  that the composite scale $m_{\nu*}$  
is the same as the excited neutrino mass $m_*$. 
In Fig.1, we show the limit of $m_*$ for 
$\lambda_W > 1$. From this figure, we find  
\bea
m_{*} > 3.4 m_W \hs{3} (\lambda_W > 1).
\ena

Firstly, we compare our new result with the old result which is 
estimated by using the effective interaction in Eq.(3). By 
comparing Eq.(28) and Eq.(5), we find that the second term 
in the parenthesis is the same as the old one, while  the 
first term is different by the cancellation among 
the poles due to $\nu^*$ and W propagator. 

Next, we compare our result with the  one by Panella et al.[8]. 
One difference is the coefficients of mass factors of 
nuclear matrix elements. They used the assumption $m_*>>m_W$. 
Our coefficients agree with them  in this limit. We evaluated 
without this assumption to see the behavior near $m_W$.  Our 
formula is valid for 
$m_* >> m_A=0.85$GeV. Another difference is 
the overall factor. Our decay rate formula is about sixteen times 
larger than their formula for $m_*>>m_W$ case where we can 
compare with their formula. As a result, we find 
the factor $2$ stringent limit on the composite scale 
$m_{\nu*}$  or the relative coupling $\lambda_W$.  These 
can be seen from Fig.2 and Fig.3. In Fig.2, we showed the 
lower bound on $\Lambda_c \equiv m_{\nu*}/\sqrt{2}$ 
for $\lambda_W>1$. The lower bound at $m_*=6m_W$ is about 
0.15 while their bound is about 0.08. In Fig.3, we showed 
the upper bound on $\lambda_W$ when $\Lambda_c 
\equiv m_{\nu*}/\sqrt{2}=$1TeV. Again our bound is about factor 
two severer than their bound.  

In summary, we computed the half-life formula of the neutrinoless 
double beta decay for the $0^+\to 0^+$ transition. We made the 
systematic analysis without assuming $m_*>>m_W$. Our result 
differs from the one by Panella et al., some mass factors 
due to their assumption $m_*>>m_W$ and the normalization 
difference about sixteen times.

\newpage
\noindent
Figure Captions:
\vskip 5mm

\noindent
Fig.1: The lower bound on the excited neutrino mass $m_*$ 
for $m_{\nu*}=m_*$ and $\lambda_W>1$. The allowed mass 
range is the region $m_*>3.4m_W$.

\vskip 5mm
\noindent
Fig.2: The lower bound on the composite scale $\Lambda_c\equiv 
m_{\nu*}/\sqrt{2}$ as a function of $m_*/m_W$. 

\vskip 5mm
\noindent
Fig.3: The upper bound on the relative coupling $\lambda_W$ 
for $\Lambda_c\equiv m_{\nu*}/\sqrt{2}>1$TeV as a function of 
$m_*/m_W$.

\end{document}